\documentclass[twocolumn,showpacs,preprintnumbers,amsmath,amssymb]{revtex4}
\usepackage{graphicx}
\usepackage{dcolumn}
\usepackage{bm}

\begin{document}

\preprint{APS/123-QED}

\title{Comment on ``Quantum key distribution for $d$-level systems with generalized Bell states" [Phys. Rev. A 65, 052331 (2002)]}

\author{Fei Gao$^{1,2}$, \quad Fenzhuo Guo$^{1}$, \quad Qiaoyan Wen$^{1}$, and Fuchen Zhu$^{3}$\\
        (1. School of Science, Beijing University of Posts and Telecommunications, Beijing, 100876, China) \\
        (2. State Key Laboratory of Integrated Services Network, Xidian University, Xi'an, 710071, China)\\
        (3. National Laboratory for Modern Communications, P.O.Box 810, Chengdu, 610041, China)\\ Email: hzpe@sohu.com}

\date{\today}

\begin{abstract}
In the paper [Phys. Rev. A \textbf{65}, 052331(2002)], an
entanglement-based quantum key distribution protocol for $d$-level
systems was proposed. However, in this Comment, it is shown that
this protocol is insecure for a special attack strategy.
\end{abstract}

\maketitle

In the paper \cite{KB}, V. Kariminpour \textit{et al}. presented a
quantum key distribution (QKD) protocol for $d$-level systems
based on shared entanglement of a reusable Bell state. The
security against some individual attacks is proved, where the
information gain of Eve is zero and the QBER introduced by her
intervention is $(d-1)/d$. However, in this paper we will show
that, by a special attack strategy Eve can get about half of the
key dits without being detected by Alice and Bob.

\begin{figure}
\includegraphics{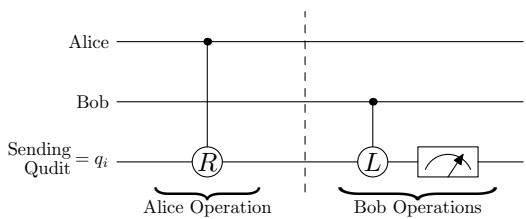}
\caption{\label{fig:one} The QKD protocol. Note that in this
Comment, for simplicity, the operation $H\otimes H^{\ast}$ or
$H\otimes H^{\ast}\otimes H$ is not included in our figures.}
\end{figure}

For convenience, we use the same notations as in Ref.\cite{KB}.
Let us give a brief description of the QKD protocol firstly (see
Fig.~\ref{fig:one}). At the beginning, Alice and Bob share a
generalized Bell state
\begin{equation}
|\Psi_{00}\rangle=\frac{1}{\sqrt{d}}\sum_{j=0}^{d-1}|j,j\rangle_{a,b}.
\end{equation}
Denote the $i$-th key dit to be sent by $q_i$, which is encoded as
a basis state $|q_i\rangle_k$. To send the key dit $q_i$ to Bob,
Alice performs a controlled-right shift on $|q_i\rangle_k$ and
thus entangles this qudit to the previously shared Bell state.
Then she transmits this qudit to Bob, from which Bob can obtain
the key dit $q_i$ by performing a controlled-left shift and a
measurement on it. Because every sending qudit is in a completely
mixed state, Eve can not extract information about the key.
Furthermore, to strengthen the security of this protocol, Alice
and Bob perform $H\otimes H^{\ast}$ on their Bell states before
encrypt each $|q_i\rangle_k$.

\begin{figure}
\includegraphics{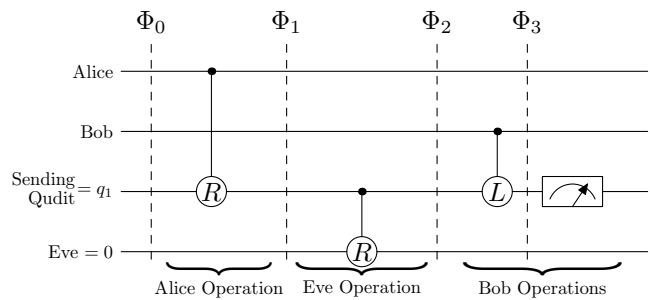}
\caption{\label{fig:two} Eve's attack in the first round.}
\end{figure}

We will describe Eve's strategy separately for each qudit.
Hereafter we use the term ``the $i$-th round" to denote the
processing procedures of the $i$-th qudit, and Alice and Bob's
operation $H\otimes H^{\ast}$ is taken as the beginning of each
round. In addition, we use $|\psi_{i0}\rangle_{a,b,e}$ and
$|\psi_{i1}\rangle_{a,b,e}$ to denote the states shared by Alice,
Bob and Eve in the beginning and the end of the $i$-th round,
respectively. Suppose Eve prepares $|0\rangle$ as her ancilla, the
eavesdropping strategy can be described as follows:

(i) In the first round, Eve entangles her ancilla into the Bell
state shared by Alice and Bob. More specifically, Eve intercepts
the sending qudit and performs a controlled-right shift on her
ancilla, then resends the sending qudit to Bob (see
Fig.~\ref{fig:two}). The initial state of Alice, Bob and Eve's
particles can be represented as
\begin{eqnarray}
|\psi_{10}\rangle_{a,b,e}=\frac{1}{\sqrt{d}}\sum_{j=0}^{d-1}|j,j,0\rangle_{a,b,e}.
\end{eqnarray}
Then the states at various stages in Fig.~\ref{fig:two} are as
follows:
\begin{eqnarray}
|\Phi_0\rangle&=&\frac{1}{\sqrt{d}}\sum_{j=0}^{d-1}|j,j,q_1,0\rangle_{a,b,k,e},\\
|\Phi_1\rangle&=&\frac{1}{\sqrt{d}}\sum_{j=0}^{d-1}|j,j,j+q_1,0\rangle_{a,b,k,e},\\
|\Phi_2\rangle&=&\frac{1}{\sqrt{d}}\sum_{j=0}^{d-1}|j,j,j+q_1,j+q_1\rangle_{a,b,k,e},
\end{eqnarray}
\begin{eqnarray}
|\Phi_3\rangle&=&\frac{1}{\sqrt{d}}\sum_{j=0}^{d-1}|j,j,q_1,j+q_1\rangle_{a,b,k,e}.
\end{eqnarray}

In the last stage, when Bob performs his controlled-left shift, he
disentangles the key qudit $|q_1\rangle_k$ and correctly gets the
value of $q_1$, while the original Bell state has now been
entangled with the state of Eve in the form of
\begin{equation}
|\psi_{11}\rangle_{a,b,e}=\frac{1}{\sqrt{d}}\sum_{j=0}^{d-1}|j,j,j+q_1\rangle_{a,b,e}.
\end{equation}

(ii) In the second round, Eve tries to avoid the detection and, at
the same time, retain her entanglement with Alice and Bob. As was
proved in Ref.\cite{KB}, Eve can not obtain information in this
round. However, we will show that she can take some measures to
avoid the detection.

Firstly, when Alice and Bob perform the operations $H\otimes
H^{\ast}$ on their ``Bell state", Eve also performs $H$ on her
ancilla. As a result, the entangled state of Alice, Bob and Eve
will be converted into
\begin{eqnarray}
|\psi_{20}\rangle_{a,b,e}&=&H\otimes H^{\ast}\otimes H |\psi_{11}\rangle_{a,b,e}\nonumber\\
&=&\frac{1}{\sqrt{d}}\sum_{j=0}^{d-1}H\otimes H^*\otimes H|j,j,j+q_1\rangle_{a,b,e}\nonumber\nonumber\\
&=&\frac{1}{d^2}\sum_{j,k,l,m=0}^{d-1}\zeta^{jk-jl+m(j+q_1)}|k,l,m\rangle_{a,b,e}.
\end{eqnarray}
Summing over $j$ and using the identity
$\frac{1}{d}\sum_{j=0}^{d-1}\zeta^{jn}=\delta(n,0)$, we finally
arrive at
\begin{equation}
|\psi_{20}\rangle_{a,b,e}=\frac{1}{d}\sum_{k,l=0}^{d-1}\zeta^{q_1(l-k)}|k,l,l-k\rangle_{a,b,e}.
\end{equation}
\begin{figure}
\includegraphics{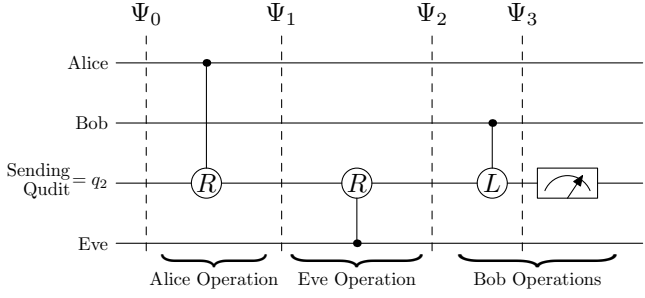}
\caption{\label{fig:three} Eve's attack in the second round.}
\end{figure}

Afterwards, Eve intercepts the sending qudit, performs a
controlled-right shift on it, and then resends it to Bob (see
Fig.~\ref{fig:three}). The states at various stages in
Fig.~\ref{fig:three} are as follows:
\begin{eqnarray}
|\Psi_0\rangle&=&\frac{1}{d}\sum_{k,l=0}^{d-1}\zeta^{q_1(l-k)}|k,l,q_2,l-k\rangle_{a,b,k,e},
\end{eqnarray}
\begin{eqnarray}
|\Psi_1\rangle&=&\frac{1}{d}\sum_{k,l=0}^{d-1}\zeta^{q_1(l-k)}|k,l,k+q_2,l-k\rangle_{a,b,k,e},\\
|\Psi_2\rangle&=&\frac{1}{d}\sum_{k,l=0}^{d-1}\zeta^{q_1(l-k)}|k,l,l+q_2,l-k\rangle_{a,b,k,e},\\
|\Psi_3\rangle&=&\frac{1}{d}\sum_{k,l=0}^{d-1}\zeta^{q_1(l-k)}|k,l,q_2,l-k\rangle_{a,b,k,e}.
\end{eqnarray}

In the last stage, when Bob performs his controlled-left shift, he
disentangles the key qudit $|q_2\rangle_k$ and correctly gets the
value of $q_2$, while leaving the state
\begin{eqnarray}
|\psi_{21}\rangle_{a,b,e}=\frac{1}{d}\sum_{k,l=0}^{d-1}\zeta^{q_1(l-k)}|k,l,l-k\rangle_{a,b,e}.
\end{eqnarray}

(iii) In the third round, Eve eavesdrops the key qudit. Firstly,
as in step.(ii), Eve also performs $H$ on her ancilla when Alice
and Bob perform $H$ and $H^\ast$ on their respective particles.
The entangled state will be changed into
\begin{eqnarray}
|\psi_{30}\rangle_{a,b,e}&=&H\otimes H^*\otimes
H|\psi_{21}\rangle_{a,b,e}\nonumber\\
&=&\frac{1}{\sqrt{d}}\sum_{m=0}^{d-1}|m,m,m-q_1\rangle_{a,b,e}.
\end{eqnarray}
\begin{figure}
\includegraphics{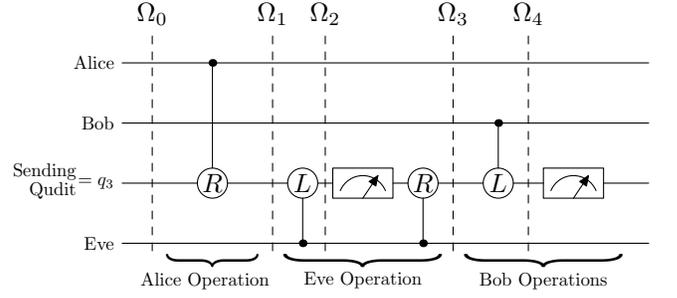}
\caption{\label{fig:four} Eve's attack in the third round.}
\end{figure}

Afterwards, Eve intercepts the sending qudit, performs a
controlled-left shift, a measurement and a controlled-right shift
on it, and then resends it to Bob (see Fig.~\ref{fig:four}). The
states at various stages in Fig.~\ref{fig:four} are as follows:
\begin{eqnarray}
|\Omega_0\rangle&=&\frac{1}{\sqrt{d}}\sum_{m=0}^{d-1}|m,m,q_3,m-q_1\rangle_{a,b,k,e},\\
|\Omega_1\rangle&=&\frac{1}{\sqrt{d}}\sum_{m=0}^{d-1}|m,m,m+q_3,m-q_1\rangle_{a,b,k,e},\\
|\Omega_2\rangle&=&\frac{1}{\sqrt{d}}\sum_{m=0}^{d-1}|m,m,q_3+q_1,m-q_1\rangle_{a,b,k,e},\\
|\Omega_3\rangle&=&\frac{1}{\sqrt{d}}\sum_{m=0}^{d-1}|m,m,m+q_3,m-q_1\rangle_{a,b,k,e},
\end{eqnarray}
\begin{eqnarray}
|\Omega_4\rangle&=&\frac{1}{\sqrt{d}}\sum_{m=0}^{d-1}|m,m,q_3,m-q_1\rangle_{a,b,k,e}.
\end{eqnarray}

It can be seen that Eve disentangles the key qudit by a
controlled-left shift, performs a measurement, and then restores
the entangled state by a controlled-right shift. As a result, Eve
obtains the measurement result $q_3+q_1$ and Bob correctly gets
the value of $q_3$. At last, the entangled state of Alice, Bob and
Eve can be written as
\begin{eqnarray}
|\psi_{31}\rangle_{a,b,e}=\frac{1}{\sqrt{d}}\sum_{m=0}^{d-1}|m,m,m-q_1\rangle_{a,b,e}.
\end{eqnarray}

(iv) In the fourth round, Eve uses the same strategy as in the
second round to avoid the detection, that is, the strategy in
step.(ii). After their operation $H\otimes H^{\ast}\otimes H$,
Alice, Bob and Eve change the entangled state into
\begin{eqnarray}
|\psi_{40}\rangle_{a,b,e}&=&H\otimes H^*\otimes H|\psi_{31}\rangle_{a,b,e}\nonumber\\
&=&\frac{1}{d}\sum_{k,l=0}^{d-1}\zeta^{-q_1(l-k)}|k,l,l-k\rangle_{a,b,e}.
\end{eqnarray}

Then Eve performs the operations as described in
Fig.~\ref{fig:three}. The states at various stages are as follows:
\begin{eqnarray}
|\Theta_0\rangle&=&\frac{1}{d}\sum_{k,l=0}^{d-1}\zeta^{-q_1(l-k)}|k,l,q_4,l-k\rangle_{a,b,k,e},\\
|\Theta_1\rangle&=&\frac{1}{d}\sum_{k,l=0}^{d-1}\zeta^{-q_1(l-k)}|k,l,k+q_4,l-k\rangle_{a,b,k,e},\\
|\Theta_2\rangle&=&\frac{1}{d}\sum_{k,l=0}^{d-1}\zeta^{-q_1(l-k)}|k,l,l+q_4,l-k\rangle_{a,b,k,e},\\
|\Theta_3\rangle&=&\frac{1}{d}\sum_{k,l=0}^{d-1}\zeta^{-q_1(l-k)}|k,l,q_4,l-k\rangle_{a,b,k,e},
\end{eqnarray}
where $\Theta_p$ corresponds to the state $\Psi_p$ in
Fig.~\ref{fig:three} ($p=0,1,2,3$).

It can be seen that, in the last stage, Bob correctly gets the
value of $q_4$, while leaving the state
\begin{eqnarray}
|\psi_{41}\rangle_{a,b,e}=\frac{1}{d}\sum_{k,l=0}^{d-1}\zeta^{-q_1(l-k)}|k,l,l-k\rangle_{a,b,e}.
\end{eqnarray}

(v) In the fifth round, Eve uses the same strategy as in the third
round to eavesdrop the key qudit, that is, the strategy in
step.(iii). After their operation $H\otimes H^{\ast}\otimes H$,
Alice, Bob and Eve change the entangled state into
\begin{eqnarray}
|\psi_{50}\rangle_{a,b,e}&=&H\otimes H^*\otimes H|\psi_{41}\rangle_{a,b,e}\nonumber\\
&=&\frac{1}{\sqrt{d}}\sum_{j=0}^{d-1}|j,j,j+q_1\rangle_{a,b,e}.
\end{eqnarray}

Then Eve performs the operations as described in
Fig.~\ref{fig:four}. The states at various stages are as follows:
\begin{eqnarray}
|\Upsilon_0\rangle&=&\frac{1}{\sqrt{d}}\sum_{j=0}^{d-1}|j,j,q_5,j+q_1\rangle_{a,b,k,e},\\
|\Upsilon_1\rangle&=&\frac{1}{\sqrt{d}}\sum_{j=0}^{d-1}|j,j,j+q_5,j+q_1\rangle_{a,b,k,e},\\
|\Upsilon_2\rangle&=&\frac{1}{\sqrt{d}}\sum_{j=0}^{d-1}|j,j,q_5-q_1,j+q_1\rangle_{a,b,k,e},\\
|\Upsilon_3\rangle&=&\frac{1}{\sqrt{d}}\sum_{j=0}^{d-1}|j,j,j+q_5,j+q_1\rangle_{a,b,k,e},\\
|\Upsilon_4\rangle&=&\frac{1}{\sqrt{d}}\sum_{j=0}^{d-1}|j,j,q_5,j+q_1\rangle_{a,b,k,e},
\end{eqnarray}
where $\Upsilon_p$ corresponds to the state $\Omega_p$ in
Fig.~\ref{fig:four} ($p=0,1,2,3,4$). It can be seen that Eve's
measurement result in this round is $q_5-q_1$.

Obviously, in the last stage, Bob correctly gets the value of
$q_5$, while leaving the state
\begin{eqnarray}
|\psi_{51}\rangle_{a,b,e}=\frac{1}{\sqrt{d}}\sum_{j=0}^{d-1}|j,j,j+q_1\rangle_{a,b,e}.
\end{eqnarray}

Note that $|\psi_{51}\rangle_{a,b,e}=|\psi_{11}\rangle_{a,b,e}$.
Therefore, in the following rounds, Eve can use the same strategy
as in the steps from (ii) to (v) repeatedly until the last key dit
were transmitted.

Now let us give a concretely description of our eavesdropping
strategy:
\begin{description}
\item [\quad 1.] In the first round, Eve performs the operations
as described in Fig.~\ref{fig:two}; \item [\quad 2.] When Alice
and Bob perform $H$ and $H^\ast$ on their respective particles at
the beginning of every round (except for the first round), Eve
also performs $H$ on her ancilla; \item [\quad 3.] In every even
round, Eve performs the operations as described in
Fig.~\ref{fig:three}; \item [\quad 4.] In every odd round (except
for the first round), Eve performs the operations as described in
Fig.~\ref{fig:four}.
\end{description}

From the above analysis, we can see that in our eavesdropping
strategy no error will be introduced to the key distribution
between Alice and Bob, and Eve will obtain exactly the result of
$$q_3+q_1,q_5-q_1,q_7+q_1,q_9-q_1,\dots$$
from which she can infer about half of the key dits by checking
$d$ possible values for $q_1$. It should be emphasized that there
is another profitable fact for Eve. That is, at the end of QKD
procedure, Alice and Bob will compare a subsequence of the key
dits publicly to detect eavesdropping, which will leak useful
information to Eve. More specifically, as long as any odd numbered
key dit is announced, Eve can determine which of the $d$ possible
results is true.

In conclusion, though Eve cannot get information about the key dit
in every even rounds (as proved in Ref.\cite{KB}), she can take
some more clever measures to avoid the detection and retain her
entanglement with Alice and Bob, so that she can eavesdrop the key
dit in the next round. Our attack strategy is exactly based on
this fact. By our strategy Eve can obtain about half of the key
dits without being detected by Alice and Bob. Consequently the QKD
protocol in Ref.\cite{KB} is insecure against this type of attack.

This work is supported by the National Natural Science Foundation
of China, Grants No. 60373059; the National Laboratory for Modern
Communications Science Foundation of China, Grants No.
51436020103DZ4001; the National Research Foundation for the
Doctoral Program of Higher Education of China, Grants No.
20040013007; and the ISN Open Foundation.

\end{document}